\documentclass[%
reprint,
superscriptaddress,
amsmath,amssymb,
aps,
prb,
floatfix
]{revtex4-2}

\usepackage{multirow}
\usepackage{graphicx}
\usepackage{dcolumn}
\usepackage{bm}
\usepackage[version=3]{mhchem}
\usepackage{nicefrac}

\usepackage{color}
\usepackage{xcolor}
\usepackage{dcolumn}
\usepackage{amssymb}
\usepackage{url}
\usepackage{hyperref}
\usepackage{xfrac}
\usepackage{tabularx}
\usepackage{xspace}
\usepackage{amsmath}
\usepackage[normalem]{ulem}
\usepackage{mathtools}
\usepackage{physics}
\usepackage{dsfont}
\usepackage{float}
\usepackage{braket}
\usepackage{placeins}
\usepackage{footmisc}

\def\be{\begin{equation}}
\def\ee{\end{equation}}
\def\bea{\begin{eqnarray}}
\def\eea{\end{eqnarray}}
\def\ba{\begin{array}}
\def\ea{\end{array}}

\begin{document}
\title{Simplex Crystal Ground State and Magnetization Plateaus in the Spin-$\mathbf{1/2}$ Heisenberg Model on the Ruby Lattice}
\author{Pratyay Ghosh}
\email{pratyay.ghosh@epfl.ch}
\affiliation{Institute of Physics, Ecole Polytechnique Fédérale de Lausanne (EPFL), CH-1015 Lausanne, Switzerland}

\author{Fr\'ed\'eric Mila}
\affiliation{Institute of Physics, Ecole Polytechnique Fédérale de Lausanne (EPFL), CH-1015 Lausanne, Switzerland}

\begin{abstract}
We investigate the spin-$1/2$ Heisenberg antiferromagnet on the ruby lattice with uniform first- and second-neighbor interactions, which forms a two-dimensional network of corner-sharing tetrahedra. Using infinite projected entangled pair states (iPEPS), we study the ground state of the system to find that it assumes a gapped threefold-degenerate simplex crystal ground state, with strong singlets formed on pairs of neighboring triangles. We argue that the formation of the simplex singlet ground state at the isotropic point relates to the weak inter-triangle coupling limit where an effective spin-chirality Hamiltonian on the honeycomb lattice exhibits an extensively degenerate ground state manifold of singlet coverings at the mean-field level. Under an applied Zeeman field, the iPEPS simulations uncover magnetization plateaus at $m/m_s = 0, 1/3, 1/2,$ and $2/3$, separated by intermediate supersolid phases, all breaking the sixfold rotational symmetry of the lattice. Unlike the checkerboard lattice, these plateaus cannot be described by strongly localized magnons.
\end{abstract}

\maketitle

\section{Introduction}
\begin{figure}[t]
    \includegraphics[width=0.95\columnwidth]{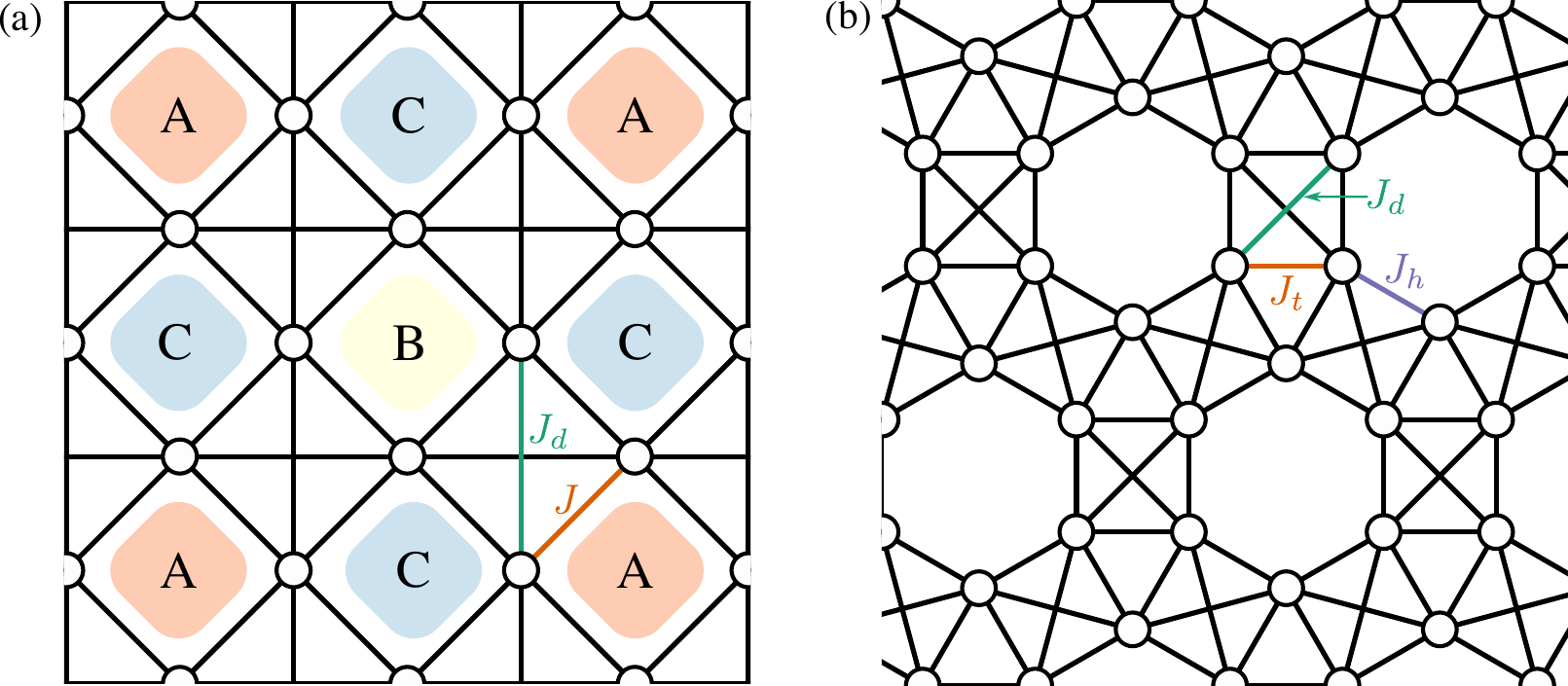}
    \caption{(a) Checkerboard lattice with first-neighbor ($J$) and second-neighbor ($J_d$) interactions. Empty plaquettes are labeled with A, B, and C, regions where localized magnon states can harbor. (b) Ruby lattice with nearest-neighbor ($J_t$ and $J_h$) and next-nearest-neighbor ($J_d$) interactions. In the isotropic interaction limit, both lattices can be viewed as networks of corner-sharing tetrahedra, but with different connectivity.} \label{fig-lattice}
\end{figure}

A central motivation for studying frustrated quantum magnets lies in their capacity to exhibit quantum disordered states and spin-liquid behavior~\cite{frustrationbook,Diepbook}. 
Among the most highly frustrated and promising candidates in this context is the pyrochlore lattice, a three-dimensional network of corner-sharing tetrahedra that has attracted substantial experimental and theoretical attention due to its unusual low-temperature behavior and the emergence of collective low-energy excitations~\cite{PhysRevLett.80.2929,RevModPhys.82.53,PhysRevX.9.011005,Ghosh2019}. 
In order to simplify the analysis of the three-dimensional pyrochlores while retaining the magnetic frustration, it is common to study the planar projection of the lattice, also known as the checkerboard lattice [see Figure \ref{fig-lattice}(a)] \cite{CB1,CB2,CB3,CB4,CB5,Richter2004_magnon,Richter2005,PhysRevB.86.041106,PhysRevB.85.134427}. 
In this mapping, each tetrahedron transforms into a square with spins at its vertices, and the six edges become the four nearest-neighbor (NN) bonds and the two diagonal bonds. 
The squares are then arranged in a corner-shared structure, as illustrated in Figure \ref{fig-lattice}(a).
In this planar version, in principle, the NN bonds can be assigned a coupling strength $J$, while the diagonal bonds can have a different coupling strength $J_d$ since they are not symmetry equivalent.
When all the Heisenberg exchange couplings are identical ($J=J_d$), the checkerboard lattice has the same local structure as the pyrochlore lattice, forming a network of corner-sharing tetrahedra.

The isotropic Heisenberg model ($J = J_d$) on the checkerboard lattice exhibits several interesting physical properties. Owing to the corner-sharing geometry, the Hamiltonian can be written as a sum over the total spin on each tetrahedron (up to an additive constant). 
Classically, this leads to a continuous ground-state degeneracy and extensive entropy. 
As a result, for $S = 1/2$, the system is known to favor a nonmagnetic state with a large spin gap~\cite{PhysRevB.64.094412,CB4,CB5}. 
A further notable feature of this model is the presence of magnetization plateaus in the field-dependent magnetization.
Plateaus occur at $m/m_s = 0,\ 1/4,\ 1/2,$ and $3/4$, where $m_s$ denotes the saturation magnetization, and each plateau corresponds to a valence bond crystal (VBC) phase~\cite{CB5,CB1,CB_mag1} whose structure can be traced to the exact localized-magnon states that appear immediately below the saturation field~\cite{CB1,Richter2004_magnon} stabilizing the $m/m_s = 3/4$ plateau. This state is an exact ground state given by a product of single-magnon localized states on (say) all the A-plaquettes together with a fully polarized (FP) configuration on the remaining spins. 
The rest of the plateaus are not exact product eigenstates, but they are well approximated by analogous localized-magnon constructions~\cite{CB1}. Specifically, the $m/m_s = 1/2$ plateau is composed of localized two-magnon states (singlets) on the A-plaquettes, and FP spins on the B-plaquettes; the $m/m_s = 1/4$ plateau contains localized two-magnon states on the A-plaquettes and single-magnon states on the B-plaquettes; and the $m/m_s = 0$ plateau is formed by localized two-magnon states on both A- and B-plaquettes.          

In the present article, we investigate another variant of a system built from corner-sharing tetrahedra, shown in Figure.~\ref{fig-lattice}(b). 
In fact, this is the only other two-dimensional lattice composed of corner-sharing tetrahedra in which the voids are regular polygons~\footnote{Each vertex is shared by two tetrahedra, and each bond (edge) belongs to one and only one tetrahedron. 
The only vertex configuration (made from uniform polygons) that satisfies these constraints are $4\cdot4^*\cdot4\cdot4^*$ and $3\cdot4^*\cdot6\cdot4^*$, where the numbers indicate the number of edges of the constituent regular polygons. For clarity, we label the tetrahedron by $4^{*}$, which functions as a square in the vertex construction. The configuration $4\cdot4^*\cdot4\cdot4^*$ yields the checkerboard lattice, and $3\cdot4^*\cdot6\cdot4^*$ corresponds to the ruby lattice with second-neighbor bonds.}.
The resulting structure is the so-called ruby lattice (or bounce lattice) with second-neighbor interactions. This lattice has attracted substantial theoretical~\cite{RLTh1,RLTh2,RlTh3,RlTh4,maity2024gappedgaplessquantumspin,Ghosh2022,Ghosh2023,Farnell2011,pac2025magneticorderingoutofplaneartificial,Ruby_Kai,Ruby_Jahromi} and experimental~\cite{Rasche2013_ruby,Pauly2015_ruby2} interest. In addition, the ruby lattice has become relevant in the context of Rydberg-atom platforms, where it provides a versatile geometry for realizing correlated quantum many-body phenomena~\cite{Ryd1,Ryd2,Ryd3,Ryd4}. 
 
The ruby lattice consists of six-site rings, similar to the pyrochlore lattice, but it is even more frustrated due to the presence of three-site plaquettes. The lattice contains two symmetry-inequivalent nearest-neighbor bonds: those forming the hexagons, assigned a Heisenberg exchange $J_h$, and those forming the triangles, assigned $J_t$. The second-neighbor bonds carry an exchange $J_d$. The spin-$1/2$ Heisenberg Hamiltonian on this lattice is given by
\be\label{eq:hmail}
H=\sum_{\langle ij \rangle_k}J_k\mathbf{S}_i\cdot\mathbf{S}_j
\ee
where $\mathbf{S}_i$ are $S=1/2$ operators and $\langle ij \rangle_k$ denotes a bond of type $k\in\{h,t,d\}$.
We investigate the ground state of the system in the presence of a magnetic field $h$, that is, the Hamiltonian in \eqref{eq:hmail} is supplemented by the Zeeman term $-h\sum_iS_i^z$. The regime of primary interest in this work is the isotropic limit $J_h = J_t = J_d$. Nonetheless, we highlight the distinction among the bond types, as it becomes important later.

We study the system using a tensor network method based on the Infinite projected entangled pair states (iPEPS)~\cite{Jiang2008,Jordan2008,Verstraete2004a,Verstraete2004b,Gu2008} to probe the ground state directly in the thermodynamic limit.
iPEPS is a variational ansatz in which the wavefunction of a two-dimensional infinite system is represented by a network of tensors, and it has successfully described various frustrated models~\cite{Jordan2008,PhysRevB.84.041108,Corboz2014,PhysRevB.85.125116,PhysRevB.88.155112,PhysRevB.94.075143,star_ghosh}. 
Our iPEPS simulations reveal magnetization plateaus at $m/m_s = 0,\ 1/3,\ 1/2,$ and $2/3$, each exhibiting a symmetry-broken structure. These plateaus, however, are more complex than those in the checkerboard lattice and cannot be described as localized-magnon states. Notably, the $m/m_s=0$ ground state is a sixfold symmetry-breaking simplex crystal with a eighteen-site unit cell. Further insight can be gained by considering a deformed Hamiltonian with $J_t \gg J_h = J_d$. In this limit, the system reduces to an effective Hamiltonian involving spin and chirality degrees of freedom on the $J_t$ triangles, capturing the essential low-energy physics of the model.

The remainder of this article is organized as follows. In Sec.~\ref{sec:methods}, we describe the iPEPS methodology used to study the system. Thereafter, we present the ground state of the system at zero field in Sec.~\ref{sec:results_gs}, followed by the details of the resulting magnetization behavior in Sec.~\ref{sec:results}. Finally, Sec.~\ref{sec:conclusions} summarizes our main findings and outlines directions for future research. Additionally, sec.~\ref{sec:app1} discusses the effective Hamiltonian and the properties of the $m/m_s=0$ state.

\begin{figure*}[t]
    \includegraphics[width=0.9\textwidth]{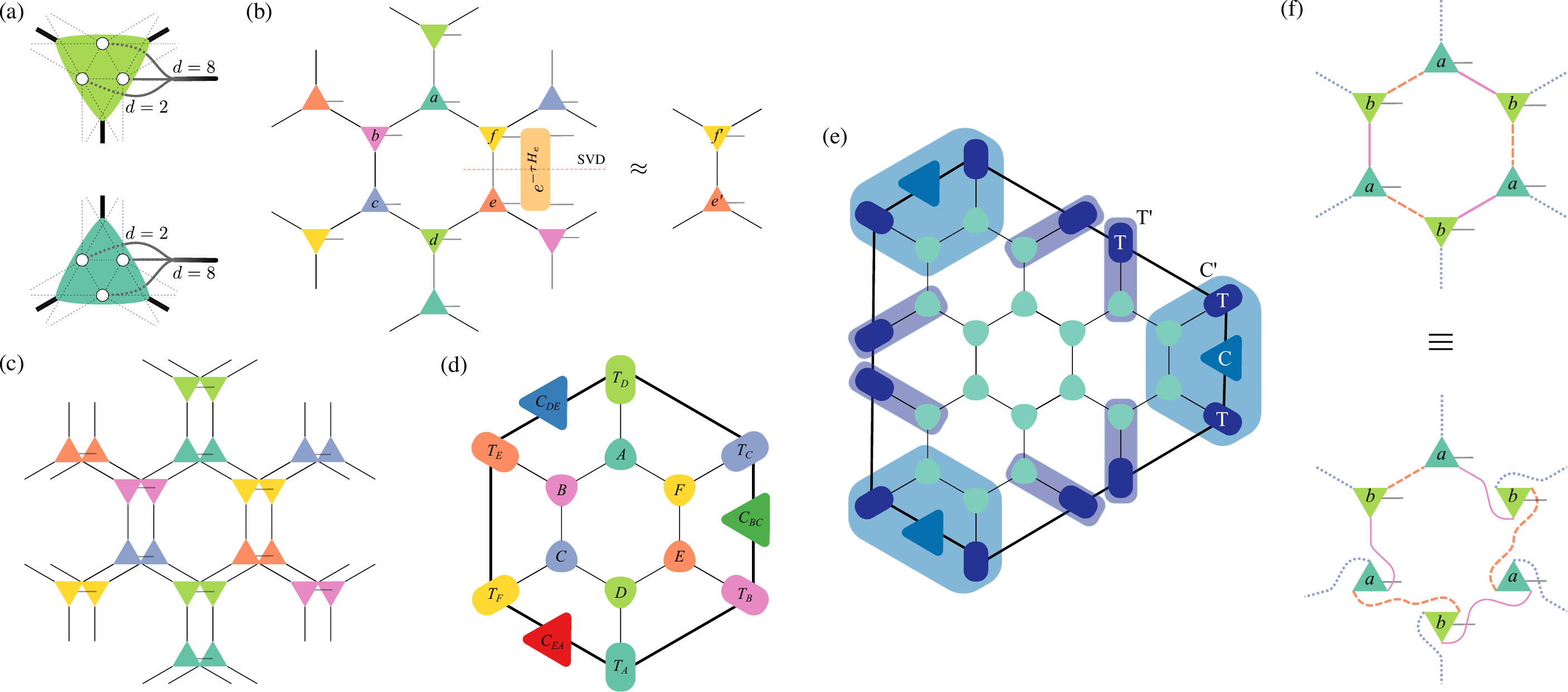}
    \caption{
    (a) Coarse-grained three-site tensors. Each $J_t$-trimer of the ruby lattice is represented by a tensor with physical dimension $d=8$. Thin dotted lines show the bonds of the underlying lattice, while thick black lines indicate virtual bonds of dimension $D$.  
    (b) 2D iPEPS wavefunction on the honeycomb network with a six-site unit cell. Black lines represent virtual legs and gray lines the physical legs. In the simple update scheme, a two-site Suzuki-Trotter gate $e^{-\tau H_e}$ is applied to each nearest-neighbor pair of tensors, followed by a singular value decomposition (SVD) across the dashed orange line separating the two physical indices. The resulting $dD^3 \times dD^2$ tensor is truncated to retain the $D$ largest singular values.  
    (c) The wavefunction overlap is represented as the contraction of the infinite tensor network. The overlapped tensor $\alpha^{[\mathbf{x}]\dagger}\alpha^{[\mathbf{x}]}$ is mapped to a local tensor $A^{[\mathbf{x}]}$ with dimensions $D^2 \times D^2 \times D^2$.  
    (d) This contraction is approximated using an environment composed of row and corner tensors; thicker black lines indicate environment bonds of dimension $\chi$.  
    (e) Schematic of the environment tensor update scheme.  
    (f) Two-trimer unit-cell setup, ordered as $a$-$b$-$R(a)$-$R(b)$-$R^2(a)$-$R^2(b)$ in the counterclockwise direction, where $R$ denotes a rotation by $2\pi/3$ about the center of the tensors.
    } \label{fig-TN}
\end{figure*}

\section{Infinite projected entangled pair states (iPEPS)}\label{sec:methods}
The iPEPS ansatz represents the ground state wavefunction of an infinite two-dimensional system as an infinite network of tensors [for example, Fig.~\ref{fig-TN}(b)]. 
In our approach, each $J_t$-trimer of the ruby lattice is represented by a single tensor with a physical leg of dimension $d = 2^3$ [see Fig.~\ref{fig-TN}(a)], resulting in a tensor-network representation of the ground state defined on a honeycomb lattice.
The tensors, denoted $\alpha_{i,j,k}^{s}(\mathbf{x})$, reside on the vertices labeled by $\mathbf{x}$ [see Fig.~\ref{fig-TN}(a)]. The indices $i,j,k$ refer to the virtual legs with dimension $D$, while the index $s$ represents the physical space.

To obtain the ground state (or a close approximation), the tensors $\alpha_{i,j,k}^{s}(\mathbf{x})$ are optimized to minimize the total energy. Several optimization strategies have been proposed for this purpose~\cite{Jiang2008,Jordan2008,Phien2015,Hasik2021}. In this work, we employ the \emph{simple update} algorithm~\cite{Jiang2008}, in which a network of random initial tensors, $|\psi_0 \rangle$, is evolved in imaginary time through a sequence of local, quasi-adiabatic updates that mimic a slow annealing process, relaxing the system towards the ground state. For gapped systems, as we will see is the case here, the entanglement is relatively short-ranged, and correlations decay quickly. In such cases, simple update is efficient and can provide an accurate approximation of the ground state as the local updates are not significantly influenced by the surrounding environment. 

The imaginary-time evolution is performed using the Trotter-Suzuki decomposition of $e^{-\tau H}$ repeatedly as 
\begin{align}
   | \psi_{GS} \rangle  & = \lim_{\beta \rightarrow \infty} e^{-\beta H} | \psi_0 \rangle \\
  & \simeq \lim_{n \rightarrow \infty} \left(\prod_{ e \in \langle \mathbf{x}, \mathbf{x'} \rangle } e^{- \tau H_{e}}\right)^n | \psi_0 \rangle \ 
\end{align}
where we set $\tau = 10^{-3}$-$10^{-2}$. The evolution is carried out by successively applying two-site gates $e^{- \tau H_{e}}$ to nearest-neighbor tensors. $H_{e}$ is the Hamiltonian acting on the spins belonging to the two tensors. Since the dimension of the local tensors $\alpha_{i,j,k}^{s}(\mathbf{x})$ would otherwise grow exponentially with the number of gates, the tensors are projected back to the relevant subspace after each gate. This is achieved via a singular value decomposition, keeping only the $D$ largest singular values, as illustrated in Fig.~\ref{fig-TN}(b).

Once the tensors have been optimized, local observables can be evaluated by contracting the infinite two-dimensional tensor network constructed from the local tensor \mbox{$\mathcal{A}_{i^{},i',j^{},j',k^{},k'}(\mathbf{x}) = \alpha_{i',j',k'}^{s}(\mathbf{x}) \alpha_{i^{},j^{},k^{}}^{s^{}}(\mathbf{x})$} [see Fig.~\ref{fig-TN}(c)]. The contraction of the infinite network is approximated using an environment comprised of row tensors $T(\mathbf{x})$ and corner tensors $C(\mathbf{x})$ [see Fig.~\ref{fig-TN}(d)] computed via the corner transfer matrix renormalisation group (CTMRG) algorithm \cite{Nishino1996,Orus2009} specifically adapted for tensor networks on the honeycomb lattice \cite{gendiar2012,nyckees2023,lukin2023,lukin2024,star_ghosh} [see Fig.~\ref{fig-TN}(e)]. The row and corner tensors have dimensions $\chi \times D^2 \times \chi$ and $\chi \times \chi$, respectively. 
The environment dimension $\chi$ acts as a control parameter, with observables expected to converge to their exact values in the $\chi \rightarrow \infty$ limit, where the results are truly variational.
It is therefore necessary to verify that the observables, such as the energy and magnetization, have converged with respect to $\chi$. In our study of the ruby lattice, we use $\chi$ up to $D(D+2)+16$, for which the energy has converged to a precision of $10^{-6}$ for all bond dimensions $D \in [3,12]$.

In our initial estimation of the ground state, we consider a unit cell composed of six trimers as seen in Fig.~\ref{fig-TN}(b). For all values of $h$, the algorithm converges reliably to the ground states which always preserve threefold rotational symmetry around the center of the hexagons. We exploit this feature to further improve our convergence by employing an alternative setup with only two tensors, as illustrated in Fig.~\ref{fig-TN}(f)~\cite{star_ghosh}~\footnote{All tensor network calculations in this work were performed using the ITensor library~\cite{itensor}.}. 

\begin{figure*}
\includegraphics[width=0.95\textwidth]{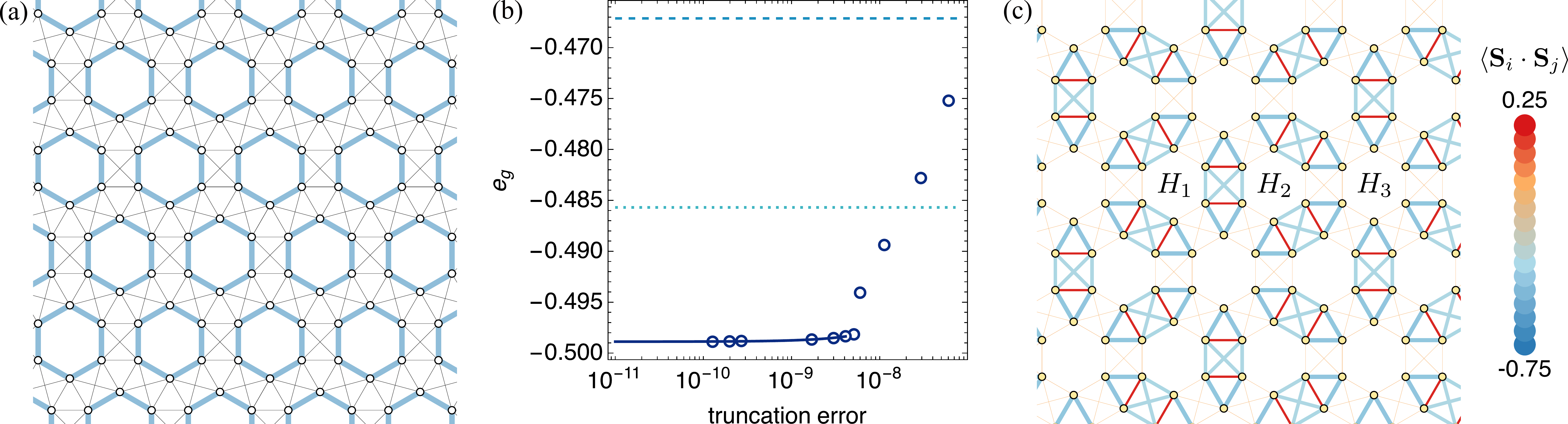}
    \caption{(a) Product state of singlets on the hexagons, a possible candidate ground state of the system. (b) Scaling of the ground state energies per site, $e_g$, obtained in iPEPS as a function of the truncation error from SVD at $J_h = J_t = J_d= 1$. The solid line shows the extrapolation for $D\to\infty$ obtained via an exponential fitting of the data obtained for $D=8$-$12$. The horizontal dashed line corresponds to the energy of the state shown in (a). The horizontal dotted line shows the per-site energy of an isolated simplex. (c) The spin-spin correlations on the first- and second-neighbor bonds of the ground state of \eqref{eq:hmail}. This state is a simplex crystal, where strong singlets form on simplexes composed of two neighboring $J_t$ triangles. The state is threefold degenerate. We mark the three inequivalent hexagonal plaquettes with $H_1$, $H_2$, and $H_3$. The legend at the extreme right indicates the spin-spin correlation values, shown as a false-color scale, and applies to both panel (a) and (c).} \label{fig-gs}
\end{figure*}
\section{Ground state for $h=0$}\label{sec:results_gs}
Drawing motivation from the plaquette singlet ground state in the uniform checkerboard lattice, one might naturally expect the ruby lattice at $J_h=J_t=J_d\ (=1)$ to favor an analogous plaquette singlet phase with dominant singlet weight on the $J_h$ hexagons. Figure~\ref{fig-gs}(a) shows a schematic representation of this state. 
This expectation, however, is not borne out on ruby lattice. The ground-state energy per spin obtained from our iPEPS calculations is $-0.498872$ [Figure~\ref{fig-gs}(b)], which is significantly lower than the energy of the hexagonal-singlet product state, $(-2-\sqrt{13})/6\approx -0.467129$~\cite{Ghosh_Hida_Model_of_kagome,Ghosh2024-al}.
We plot the spin-spin correlations of the ground state found in our iPEPS calculations for $D=12$ on the first- and second-neighbor bonds in Figure~\ref{fig-gs}(c). Visibly, this state is different from the uniform singlet state shown in Figure~\ref{fig-gs}(a). We identify the ground state of the system as a simplex crystal constructed of strong singlets on 
\begin{center}
  \includegraphics[width=0.2\columnwidth]{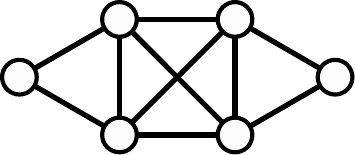}
\end{center}
simplexes. The ground state breaks the $C_6$ rotational symmetry about the hexagon centers down to $C_3$, resulting in a threefold degeneracy.
The energy per site of the Heisenberg Hamiltonian on such a simplex is $-(3 + 2 \sqrt{2})/12 = -0.485702$, which is in better agreement with the energy extracted from the iPEPS calculations than that of the hexagonal singlet product state. The excellent convergence of the energy with respect to the SVD truncation error (and with $D$), as shown in Figure~\ref{fig-gs}(b), suggests the presence of a gap. The spin gap for an isolated simplex is $0.843546$, which is also consistent with a finite gap in the ground state of the full Hamiltonian.   

One way to understand this simplex crystal ground-state follows from the limit $J_t \gg J_h=J_d$. In this region, the ruby lattice reduces to an effective spin-chirality Hamiltonian defined on a honeycomb lattice of trimer supersites, where each isolated trimer contributes a fourfold-degenerate ground-state manifold labeled by spin and chirality Pauli matrices [details in Appendix~\ref{sec:app1}]. A mean-field decoupling of the effective Hamiltonian leads to ground states that are exact dimer-product wavefunctions with extensive degeneracy equal to the number of dimer coverings of the honeycomb lattice. In the original spin language, each of these dimers translates to a simplex composed of two trimer units. This mean-field ground state is adiabatically connected to the original model at $J_h=J_d=J_t$, and its energy closely matches the iPEPS value, confirming that the ground state of the original spin Hamiltonian consists of simplex singlets stabilized by the interplay of spin and chirality degrees of freedom associated with each $J_t$ triangle. While this approach does not directly provide us with $C_3$ symmetric simplex crystal, the ordering can be understood by analogy with the quantum dimer model on the honeycomb lattice~\cite{RK,RK1}, where competing kinetic and potential terms select specific dimer patterns. Within the effective spin-chirality Hamiltonian, our ground state corresponds to the same dimer configuration realized in the columnar (also known as star) phase of the honeycomb-lattice quantum dimer model~\cite{RK,RK1}.

\begin{figure*}[t]
    \includegraphics[width=0.95\textwidth]{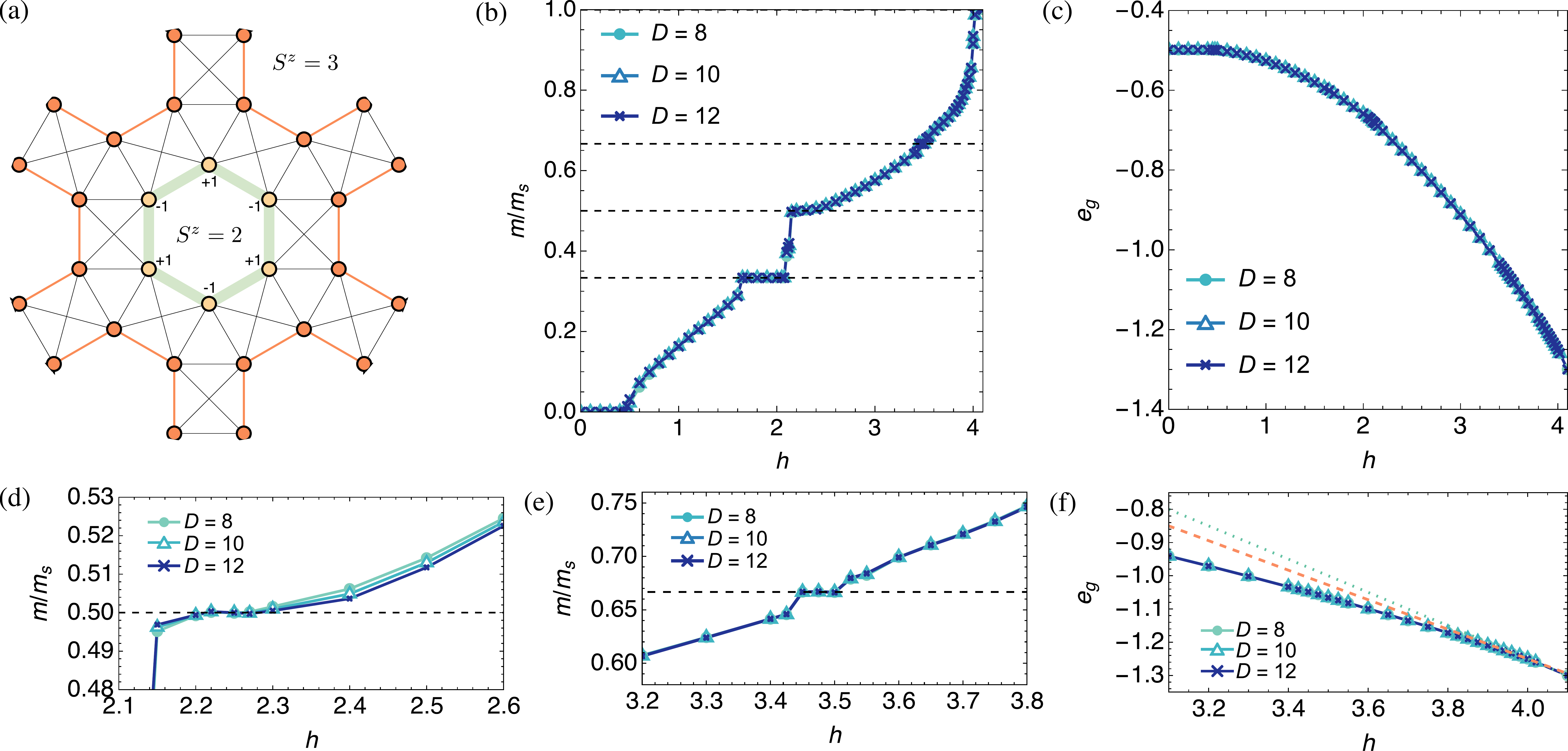}
    \caption{
    (a) Localized one magnon state on the ruby lattice. The numbers $\pm 1$ at the corners of the central hexagon denote the coefficients $c_i$ in \eqref{eq:magnon}.
    (b) Magnetization curve of \eqref{eq:hmail} with $J_t = J_h = J_d = 1$ under a Zeeman term $-h\sum_i S_i^z$. Plateaus appear at $m/m_s = 0, 1/3, 1/2,$ and $2/3$, with the stability of the $m/m_s = 0$ and $1/3$ plateaus indicating substantial spin gaps.  
    (c) Energies obtained from iPEPS calculations for various bond dimensions $D$.
    (d) and (e) Zoomed-in views of the magnetization curve near the $m/m_s = 1/2$ and $2/3$ plateaus, respectively. 
    (i) Zoomed-in view of the energies near the saturation field. The red dashed line shows the energy of the fully polarized state, while the green dotted lines correspond to localized-magnon eigenstates with $m/m_s = 2/3$ (see text).
} \label{fig-energy}
\end{figure*}

\section{Magnetization Curve and plateau states}\label{sec:results}
In the previous section, we have established that at zero field, the ground state of the system is a non-magnetic simplex crystal with a finite gap to the triplet states. One can view these triplet excitations as hard-core bosons that live on an effective kagome lattice, that is the underlying lattice of the simplexes. When a magnetic field is applied, and the spin gap closes, a number of these bosons appear, and they may form insulating and complex supersolid phases driven by nontrivial hopping structure and repulsive interactions~\cite{Ghosh2023,kagome-boson}. This behavior manifests in the magnetization curve as the plateau (solid) phases and the intermediate non-plateau regions. 
Our iPEPS simulations, which are capable of capturing plateau states with unit cells containing up to eighteen sites, yield a magnetization curve [Fig.~\ref{fig-energy}(b), (d), and (e)] featuring four magnetization plateaus at $m/m_s = 0,\ 1/3,\ 1/2,$ and $2/3$.  


Starting from the saturation field, $h_s=4$, we see that the magnetization decreases continuously until $h=3.5$, where the $m/m_s=2/3$ plateau stabilizes over a narrow region $h\in [3.45,3.5]$ as shown in Figure~\ref{fig-energy}(e). As the magnetic field is further lowered, the magnetization decreases monotonically until $h=2.3$ where the $m/m_s=1/2$ plateau emerges within $2.22<h<2.26$ [Figure~\ref{fig-energy}(d)]. Below this, the magnetization drops sharply, and the $m/m_s=1/3$ plateau stabilizes for $h\in [1.65,2.075]$ [see Figure~\ref{fig-energy}(b)]. Further reduction of the Zeeman field produces a continuous magnetization curve, with the final plateau appearing for $h\in [0,0.45]$. The existence of the $m/m_s=0$ plateau is another indication of a gapped ground state at $h=0$. 

The features immediately below $h_s$ are quite intriguing as the ruby lattice shares some geometrical similarities with the checkerboard lattice, where a sharp drop in magnetization is observed \cite{CB_mag1,CB1} followed by the stabilization of a localized-magnon state. 
One can also construct such localized-magnon states on the ruby lattice. For that, one starts by building single-magnon eigenstates ($S^z=2$) immediately below $h_s$ following the prescription of Richter \emph{et al.}~\cite{Richter2004_magnon,Richter2005,PhysRevLett.88.167207}. 
For $h>h_s$, the lowest excitation above the FP state is a single-magnon state, which is generally dispersive. On certain geometries, like our ruby lattice, however, the magnon can become strictly localized within a region $\mathcal{R}$. To illustrate this, consider a magnon wavefunction of the form
\be\label{eq:magnon}
\ket{\Phi} \sim \sum_{i \in \mathcal{R}} c_i S_i^-\ket{\text{FP}}
\ee
where $c_i$ denotes the coefficients. The remainder of the system, $\bar{\mathcal{R}}$, remains in a product state with $\ket{\Phi}$. The magnon can hop out of $\mathcal{R}$ through the exchange couplings connecting $\mathcal{R}$ and $\bar{\mathcal{R}}$. However, as there can be multiple bonds that connect a site-$j$ in $\bar{\mathcal{R}}$ to $\mathcal{R}$, one can show that destructive interference between these processes localizes the magnon when the following necessary (and sufficient in lattices with equivalent sites) condition holds:  
\be
\sum_{i\in\mathcal{R}}J_{ij} c_i = 0, \ \forall j \in \bar{\mathcal{R}}.
\ee 
On the ruby lattice, this condition is satisfied when the magnon is confined to a $J_h$ hexagon with an alternating sign structure of $c_i$, as shown in Figure \ref{fig-energy}(a) \footnote{More exotic localizations are also possible. For instance, the magnon can also localize on a one-dimensional chain built from $J_d$ bonds.}.

Following this, one can construct an exact product state of single-magnon states placed on every hexagonal plaquette, yielding $m/m_s=2/3$. This state becomes energetically favorable to the FP state at $h=4$. The energy of this state, however, is higher than the energy obtained in our iPEPS calculations [please refer to Figure~\ref{fig-energy}(f)]. Diluted variants, where single-magnon plaquettes are surrounded by FP plaquettes, can also be constructed. These states incur additional energy penalties from the bonds connecting neighboring FP plaquettes and therefore remain even higher in energy. In conclusion, the state just below the saturation field and the $m/m_s = 2/3$ plateau state obtained from the iPEPS calculations does not appear to correspond to a localized-magnon state. 

\begin{figure*}[t]
    \includegraphics[width=0.95\textwidth]{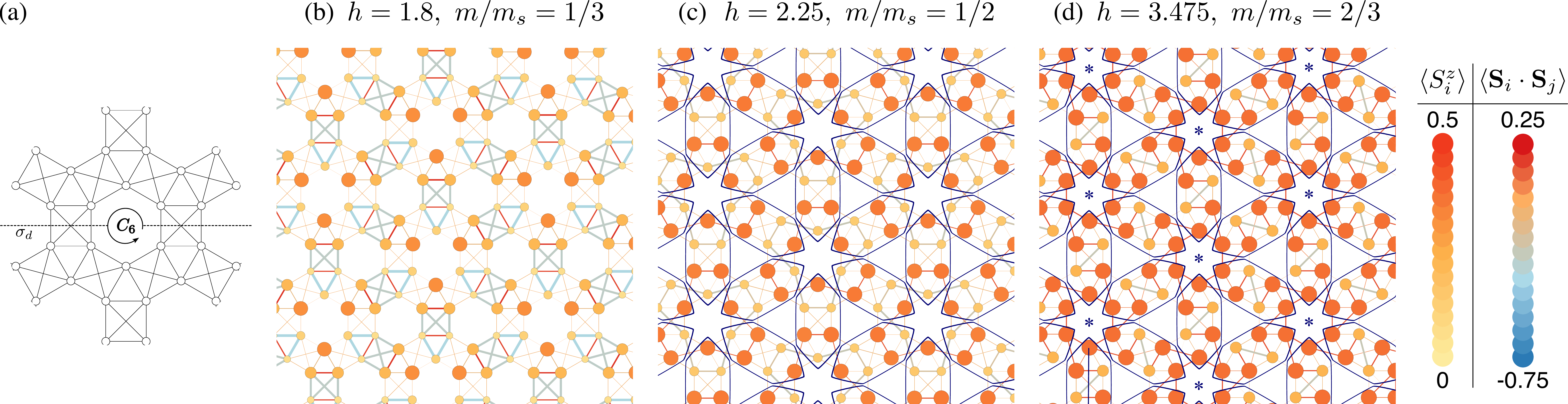}
    \caption{(a) Relevant symmetries of the ruby lattice: $C_6$ rotational symmetry through the center of the hexagons and reflection symmetry with respect to the $\sigma_d$ planes passing through opposite edges of a hexagon. (b)-(d) Magnetization and spin-spin correlations on the first- and second-neighbor bonds of the ruby lattice for $m/m_s = 1/3$, $1/2$, and $2/3$ plateau states, all of which break the $C_6$ symmetry down to $C_3$ and the reflection symmetry about the $\sigma_d$ planes, making them sixfold degenerate. The plot legends indicate the values of the spin-spin correlations and local magnetizations, represented as false-color scales. The radius of the circles is also indicative of local magnetizations. The dark-blue hexagons in panels (c) and (d) indicate the simplex patterns.
    } \label{fig-plateau}
\end{figure*}


To obtain further insights into the nature of the plateau states with $m\ne0$, we plot the magnetization profile and the spin-spin correlations for $D=12$ on the bonds in Figure~\ref{fig-plateau}(b)-(d). Visibly, all the plateau states break the sixfold rotation symmetry about the center of the hexagons down to threefold, similar to our simplex crystal singlet state. For the $m/m_s=1/3$ state, we find that each simplex carries a total $S^z = 1$. The state on these simplexes, however, is not the same uniform $S^z=1$ triplet of the Heisenberg model on an isolated simplex. This deviation results from the effective repulsion between the triplets on neighboring simplexes~\cite{Hard-Boson-SSM,Ghosh2023}. As a consequence, the system lowers its energy by redistributing the triplet weights so that the maximally polarized spins on adjacent simplexes are as far apart as possible [see Figure \ref{fig-plateau}(b)]~\cite{Corboz2014}. This leads to the breaking of an additional reflection symmetry about $\sigma_d$ [Fig.~\ref{fig-plateau}(a)], making the $m/m_s = 1/3$ state sixfold degenerate. The $m/m_s=1/2$ and $2/3$ plateau states, as seen in Figure \ref{fig-plateau}(c) and (d), also break the reflection symmetry with respect to $\sigma_d$ planes and are therefore again sixfold degenerate. A simple-minded picture for the $m/m_s=1/2$ plateau is that all the upward-pointing $J_t$ triangles host two up spins, the downward-pointing triangles contain one up spin, and the rest of the sites are in equal superposition of up and down. Similarly, the $m/m_s=2/3$ plateau corresponds to two up spins on every triangle, with the remaining sites again in up-down superpositions. It is worth noting that although we find a plateau at $m/m_s=1/2$ in the region $h \in [2.22,2.26]$, the slope of the magnetization curve for $h \gtrsim 2.26$ is very small. This raises the possibility that the state we identify as $m/m_s=1/2$ plateau is actually gapless and becomes energetically favorable at $h=2.22$, and that the apparent plateau is a numerical artifact, as our tensor network-based method is not capable of capturing a gapless state. Since the honeycomb CTMRG used here does not provide direct access to correlation lengths, we cannot supply further evidence for or against this scenario.
    
\section{Conclusions and Outlook}\label{sec:conclusions}
We investigate the magnetization behavior of the spin-$1/2$ Heisenberg antiferromagnet on the ruby lattice, including first-neighbor (coupling strength $J_t$ and $J_h$) and second-neighbor (coupling strength $J_d$) interactions. We focus on the isotropic point, $J_t=J_h=J_d$, where localized-magnon eigenstates can occur. Via the iPEPS calculations, we find that the system at zero field assumes a nonmagnetic simplex crystal ground state. Under an applied magnetic field, the system exhibits magnetization plateaus at $m/m_s=0,\ 1/3,\ 1/2$ and $2/3$, separated by intermediate supersolid phases, all of which break the $C_6$ symmetry of the lattice to $C_3$ and the $m>0$ states additionally break the reflection symmetry about the $\sigma_d$ planes. None of these states can be interpreted as strongly localized magnon states [see Figure~\ref{fig-energy}(a)]. For $m/m_s = 1/3$, each simplex carries a total $S^z = 1$, but triplet weights are redistributed to minimize repulsion between neighboring simplexes. The $m/m_s = 1/2$ and $2/3$ plateaus can be understood similarly in terms of simplex configurations. To guide the readers, we mark the simplexes with dark blue hexagons in Figure~\ref{fig-plateau}(c) and (d). 
The $m/m_s = 1/2$ plateau state consists of an alternating pattern of strongly and weakly polarized sites on each simplex, resulting in an equal number of $J_t$ triangles containing one and two strongly polarized spins. To minimize repulsion between the strongly polarized spins, the simplexes arrange such that, around each $J_h$ hexagon, the two types of $J_t$ triangles alternate, producing the pattern shown in Fig.~\ref{fig-plateau}(c). Note that this arrangement also produces an alternating pattern of weakly and strongly polarized spins on the $J_h$ hexagons. 
For the $m/m_s = 2/3$ plateau, each simplex now carries four strongly polarized sites located on the two bonds that are maximally separated within the simplex. As a result, both $J_t$ trimers of a simplex host two strongly polarized spins. For a state built from these simplexes, the density of strongly polarized sites is highest on one-third of the $J_h$ hexagons, where the apices of six simplexes meet [marked with stars in Fig.~\ref{fig-plateau}(d)]. The system then minimizes its energy by arranging the simplexes to form a pinwheel pattern of strongly polarized bonds around these hexagons, producing the plateau state shown in Fig.~\ref{fig-plateau}(d). The remaining hexagons retain an alternating pattern of weakly and strongly polarized spins. 

\begin{figure}[t]
    \includegraphics[width=0.6\columnwidth]{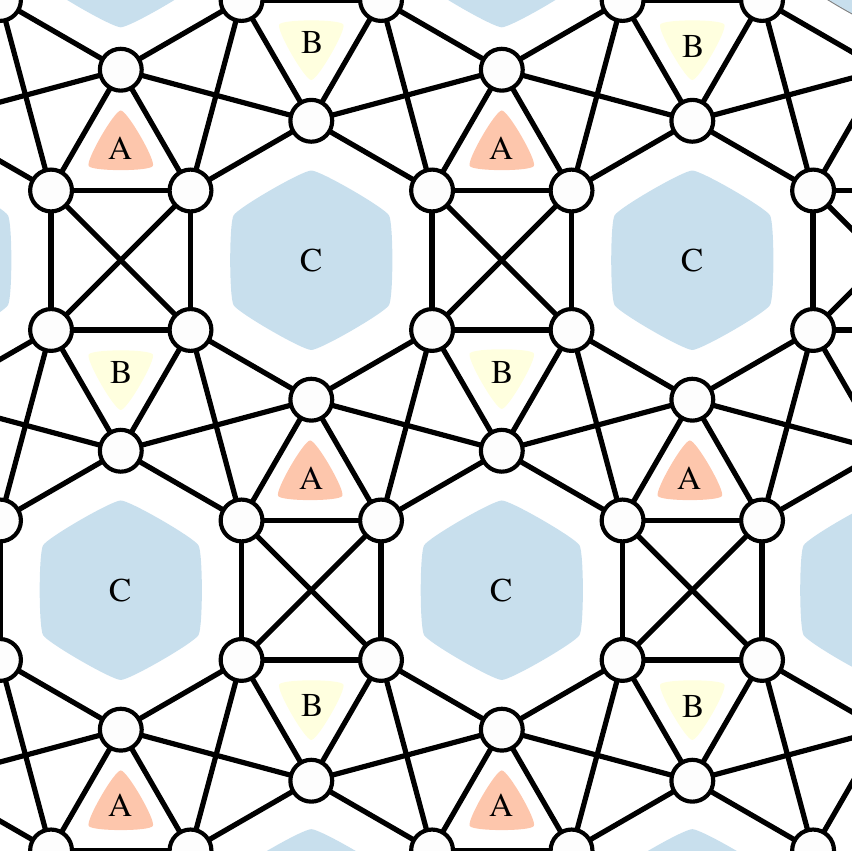}
    \caption{Ruby lattice with empty plaquettes labeled A, B, and C. At $h = 0$, the system forms strong singlets involving pairs of A and B plaquettes.} \label{fig-plaq}
\end{figure}

At first glance, the simplex crystal ground state at zero field appears to be quite different from the checkerboard lattice case, but the two are not completely unrelated. Recall that the ground state of the checkerboard lattice at $h=0$ is well approximated by plaquette singlets forming on the A and B plaquettes [see Fig.~\ref{fig-lattice}(a)]. In the Ruby lattice, we can label the plaquettes similarly: the upward-pointing $J_t$ triangles as A, the downward-pointing $J_t$ triangles as B, and the $J_h$ hexagons as C. Here, the system also tends to form singlets involving A and B plaquettes. But since A and B both are odd-site plaquettes, the singlet forms over the combined simplex instead (Figure \ref{fig-plaq}), leading to a spontaneous breaking of $C_6$ symmetry.

To understand the low-energy physics of the system at zero field, we have also studied a deformed version of the model, $J_t\gg J_h=J_d$, by performing a mean-field analysis of the effective Hamiltonian expressed in the spin and chirality basis of the isolated $J_t$ triangles. We find that the effective model enters an extensive number of singlet ground states at the mean-field level, separated from the triplet states by a substantial gap. The singlet states translate to six-site simplex coverings of the lattice. However, the iPEPS calculations find that among these states, the system selects a particular ground state with an eighteen-site unit cell and broken sixfold rotational symmetry. In our simulations, we find no indication of a hexagonal plaquette state as found in the mean-field calculations in Ref.~\cite{maity2024gappedgaplessquantumspin}, nor any other gapped spin liquid state. Our iPEPS calculations employ unconstrained tensors, which are fully capable of capturing such states~\cite{PhysRevB.86.115108,PhysRevLett.111.037202}, and would reveal them through symmetry-preserving spin-spin correlations. Notably, similar spontaneous lattice-symmetry breaking has been reported in the transverse-field Ising model on the ruby lattice, driven by an order-by-disorder mechanism~\cite{Ruby_Kai}.

The stabilization of the symmetry-breaking simplex crystal ground state is analogous to the columnar dimer pattern of the honeycomb-lattice quantum dimer model~\cite{RK,RK1,Ruby_Kai}. A similar symmetry-broken ground state has also been observed recently in the Heisenberg model on the star lattice, a system of coupled triangles closely related to the ruby lattice~\cite{star_ghosh}. Perturbative series expansions have shown that a subtle lifting of degeneracy occurs at the sixth order in inter-triangle couplings, leading the systems to choose a $C_6$ symmetry-breaking VBC state. We expect a similar mechanism to operate in the ruby lattice. In fact, exploring the phase diagram of the Heisenberg Hamiltonian as a function of $J_h = J_d$ is an interesting direction for future work. Additionally, investigating the thermal phase transitions associated with the threefold-degenerate simplex crystal state is another promising avenue. These studies, however, lie beyond the scope of the present article and will be addressed elsewhere.

\textit{Acknowledgments:}
This work is dedicated to the memory of Johannes Richter, whose studies on the localized magnon states~\cite{Richter2004_magnon,Richter2005} are a major inspiration for this work. 
The work is supported by the Swiss National Science Foundation Grant No. 212082.
Numerical computations were performed using the facilities of the Scientific IT and Application Support Center of EPFL (SCITAS).

The iPEPS data corresponding to Figs.~\ref{fig-plateau} are openly available \cite{Ghosh2025_data}.


%

\appendix
\section{Effective spin-chirality Hamiltonian for $h=0$}\label{sec:app1}
Our iPEPS calculations show that the ground state of \eqref{eq:hmail} is a $C_6$ symmetry-breaking simplex crystal composed of two neighboring $J_t$-triangles. An isolated trimer has a fourfold-degenerate ground state arising from two spin doublets combined with two possible chiralities, giving one state for each spin-chirality combination. This local degeneracy suggests that there can exist a dense set of low-lying states, from which the system selects a particular ordered configuration. To grasp the relevant low-energy physics, it is therefore natural to analyze the system in the limit $J_t \gg J_h = J_d$.

The first step is to derive an effective Hamiltonian in the subspace of the ground states of the $J_t$ triangles, as done in block-spin perturbation approach on related systems \cite{Block_Spin,kagome_mila}. We write the four ground states of an isolated triangle with two Pauli matrices: $\vec{\sigma}$ for the spin-doublet (the eigenstates of $\sigma^z$ being $\pm 1$), and $\vec{\tau}$ for its chirality (the eigenstates of $\tau^z$, denoted as $\mu$, being $\pm 1$). In the $|\sigma^z,\tau^z\rangle$ eigenbasis, the states take the form
\be
|2m,\mu\rangle =\frac{1}{\sqrt{3}}\left(|\overline{m}mm\rangle+\omega^\nu|m\overline{m}m\rangle+\omega^{\overline{\nu}}|mm\overline{m}\rangle\right)
\ee
where $\omega=\exp(i2\pi/3)$ and $|m_1m_2m_3\rangle$ represents the original spin states within one triangle. Throughout this section $\overline{x}\equiv -x$. Each triangle has energy $-3J_t/4$, and we measure all energies relative to this reference.

\begin{figure}[]
    \includegraphics[width=0.8\columnwidth]{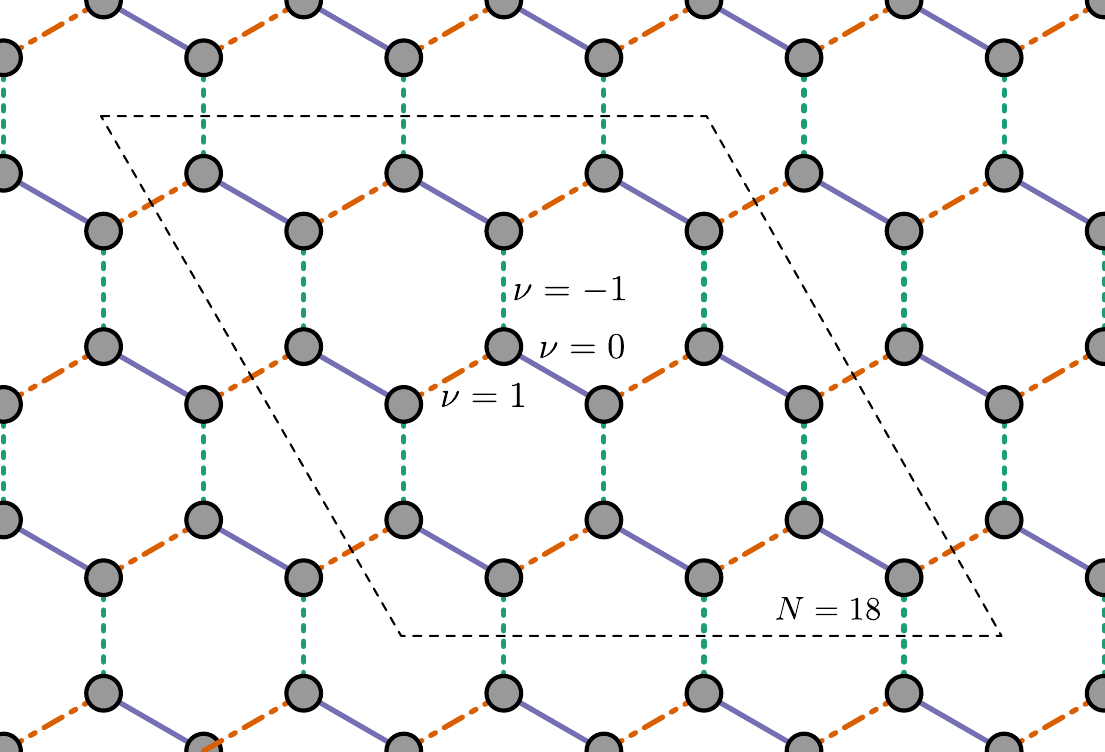}
    \caption{The effective spin-chirality Hamiltonian is defined on a honeycomb lattice with bond-dependent orbital interactions given in \eqref{eq:orb}. The cluster enclosed by the dashed line is used for the numerical solution of \eqref{eq:Hamil-mf}.} \label{fig-honeycomb}
\end{figure}

In the spin-chirality basis, it is straightforward to show that, to the first order in $J_h=J_d\equiv J'$, the ruby-lattice Hamiltonian reduces to an effective model on the   
honeycomb lattice of supersites. The resulting Hamiltonian is
\be\label{eq:eff-hamil}
\tilde{H}=\frac{J}{18}\sum_{\langle i'j'\rangle}\tilde{H}_{i'j'}^\sigma\otimes\tilde{H}_{i'j'}^\tau
\ee
with
\begin{subequations}
\be
\tilde{H}_{i'j'}^\sigma = \vec{\sigma_{i'}}.\vec{\sigma_{j'}},
\ee 
\be\label{eq:orb}
\tilde{H}_{i'j'}^\tau = \left(1+\omega^\nu \tau_{i'}^++\omega^{\overline{\nu}} \tau_{i'}^-\right)\left(1+\omega^\nu \tau_{j'}^++\omega^{\overline{\nu}} \tau_{j'}^-\right)
\ee
\end{subequations}
where $\langle i'j'\rangle$ runs over nearest-neighbor supersites, and $\nu\in\{0,\pm1\}$. For our labeling convention of the original ruby-lattice spins,  
\begin{center}
  \includegraphics[width=0.6\columnwidth]{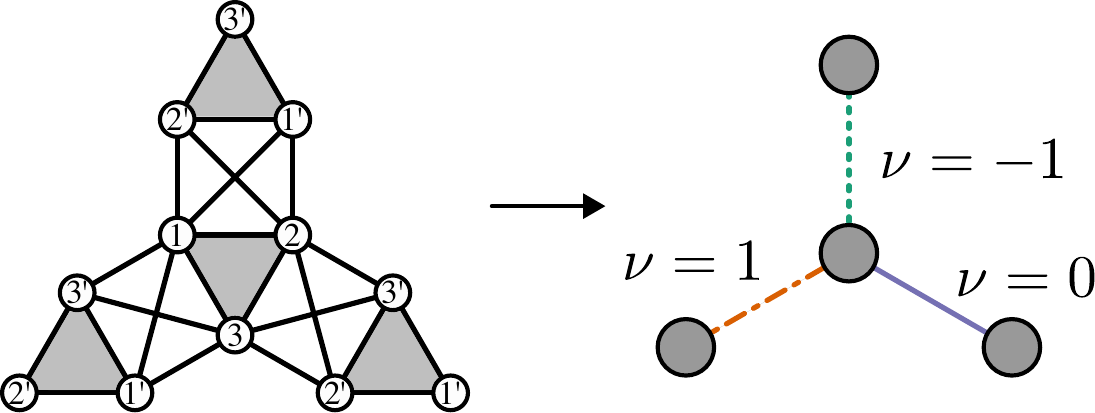},
\end{center}
resulting in three bonds emanating from each supersite to carry distinct values of $\nu$. The full pattern of bond-dependent interactions is shown in Fig.~\ref{fig-honeycomb}. \footnote{A straightforward calculation shows that $\tilde{H}_{i'j'}^\tau$ corresponds to a compass model defined on the honeycomb lattice, whose ground state is a long-range ordered dimer state~\cite{Zou2016}.}

In the chirality basis of $|\mu_{i'},\mu_{j'}\rangle=|\pm1,\pm1\rangle$, the eigenpairs of $\tilde{H}_{i'j'}^\tau$ are given by
\be
\begin{array}{rcl}
\ket{\phi_0^\tau} &=& \frac{1}{2}\left(\omega^{\overline{\nu}}, \omega^\nu, \omega^\nu, 1\right), \quad E_0^\tau = 9,\\[1ex]
\ket{\phi_1^\tau} &=& \frac{1}{2}\left(-\omega^{\overline{\nu}}, -\omega^\nu, \omega^\nu, 1\right), \quad E_1^\tau = -3,\\[1ex]
\ket{\phi_2^\tau} &=& \frac{1}{2}\left(-\omega^{\overline{\nu}}, \omega^\nu, -\omega^\nu, 1\right), \quad E_2^\tau = -3,\\[1ex]
\ket{\phi_3^\tau} &=& \frac{1}{2}\left(-\omega^{\overline{\nu}}, \omega^\nu, \omega^\nu, -1\right), \quad E_3^\tau = 1.
\end{array}
\ee
For the spin sector, $\tilde{H}_{i'j'}^{\sigma}$ has a singlet $\ket{\phi_0^{\sigma}}$ with energy $-3$ and a triplet $\ket{\phi_1^{\sigma}}$ with energy $1$.

Note that the Hamiltonian in \eqref{eq:eff-hamil} is similar to the Kugel-Khomskii model~\cite{Kugel__1982}, and it is known that when the asymmetry between spin and orbital (chirality in our case) degrees of freedom is strong, like in our case, a mean-field decoupling provides an accurate description~\cite{kagome_mila}. Accordingly, we perform a mean-field decoupling of $\tilde{H}_{i'j'}$, yielding
\be\label{eq:Hamil-mf}
\tilde{H}_{\text{MF}} = \frac{J}{18}\sum_{\langle i'j'\rangle}\left(t_{i'j'}\tilde{H}_{i'j'}^\sigma+s_{i'j'}\tilde{H}_{i'j'}^\tau -  s_{i'j'} t_{i'j'}\right)
\ee
where the mean-field parameters are $s_{i'j'}=\braket{\tilde{H}_{i'j'}^\sigma}$ and $t_{i'j'}=\braket{\tilde{H}_{i'j'}^\tau}$. The low-energy solutions can be determined analytically. For an even-site cluster with periodic boundary conditions, one can construct a wavefunction 
\be
\ket{\phi_0} = \prod_{\langle i'j'\rangle\in\mathcal{D}} \ket{\phi_0^\tau}_{i'j'}\ket{\phi_0^\sigma}_{i'j'}. 
\ee
where $\mathcal{D}$ denotes a particular covering of nearest-neighbor pairs. Following the procedure as in Ref.~\cite{kagome_mila}, one can see that $\ket{\phi_0}$ is an eigenstate of the Hamiltonian iff $s_{i'j'} = t_{i'j'} = 0$ for $\langle i'j'\rangle\notin\mathcal{D}$. For the bonds in $\mathcal{D}$, $s_{i'j'} = -3$ and $t_{i'j'} = 9$, giving an energy $E_0(S=0)=-(3J'/4)N_t$, where $N_t$ denotes the number of trimers (i.e., the number of sites in the coarse-grained honeycomb lattice). This state is a singlet, and we have verified that it is the ground state by numerically solving $\tilde{H}_{\text{MF}}$ on an 18-site cluster (marked in Fig.~\ref{fig-honeycomb}) while iteratively determining the mean-field parameters. 

The energy of the state $\ket{\phi_0}$ does not depend on the particular dimer covering $\mathcal{D}$ of the honeycomb lattice. So for a given cluster, the ground state is extensively degenerate with the degeneracy equal to the number of dimer coverings on the cluster, i.e. $1.11368^{N_t}$~\cite{PhysRev.79.357,WU2006}. In the original spin language, each dimer translates to a simplex composed of two trimer units.

For the record, the mean-field approach also motivates the construction of a few excitations. First, one can select two sites $k'$ and $l'$ which are not NNs and construct a dimer covering, $\mathcal{\bar{D}}(k',l')$, over the remaining sites and consider a wavefunction 
\be
\ket{\phi_1} = \prod_{\langle i'j'\rangle\in\mathcal{\bar{D}}(k',l')} \ket{\phi_0^\tau}_{i'j'}\ket{\phi_0^\sigma}_{i'j'} \ket{\sigma_{k'}\tau_{k'}}\ket{\sigma_{l'}\tau_{l'}}. 
\ee
where $\ket{\sigma_k'\tau_k'}$ and $\ket{\sigma_l'\tau_l}'$ can be any configuration on the unpaired sites $k'$ and $l'$. This is again a solution to $\tilde{H}_{\text{MF}}$ with energy $E_0(S=0)+3J'/2$. Each unpaired site can be interpreted as a spinon. Next, one can assume that $k'$ and $l'$ are NNs, and the spinons on them form a triplet. The corresponding wavefunction is given by
\be
\ket{\phi_2} = \prod_{\langle i'j'\rangle\in\mathcal{\bar{D}}(k',l')} \ket{\phi_0^\tau}_{i'j'}\ket{\phi_0^\sigma}_{i'j'}\ket{\phi_p^\tau}_{k'l'}\ket{\phi_1^\sigma}_{k'l'}. 
\ee
where $p=1$ or $2$. This state is also an eigenstate of $\tilde{H}_{\text{MF}}$ and has an energy $E_0(S=0)+4J'/3$, indicating that the mean-field approach predicts spinons form triplet bound states on neighboring sites with a binding energy of $J'/6$. 

We now return the original problem where $J'=J_t$. Analyzing the eigenspectrum of the two-triangle problem, we find that the hierarchy of the eigenstates remains essentially intact as $J'$ is increased from $J'\ll J_t$ to $J'=J_t$; in particular, higher-lying singlets do not move down in energy. Following the arguments of Ref.~\cite{kagome_mila}, this dictates us to expect that the ground state of the original Hamiltonian at $J'=J_t$ is adiabatically connected to the ground state at $J'\ll J_t$. For $J_h=J_d=J_t=1$, the mean-field ground state $\ket{\phi_0}$ has an energy of $-0.5$, which is remarkably close to the ground state energy of $-0.498872$ obtained from the iPEPS calculations. It is worth noting that the mean-field state is not variational, so it is not surprising that its energy is slightly lower than the iPEPS result.

\end{document}